\documentclass[twocolumn]{aastex631}

\shorttitle{Reflection spectra of super-Eddington X-ray sources}
\shortauthors{S. Shashank et al.}

\begin{document}

\title{Modeling reflection spectra of super-Eddington X-ray sources}

\author[0000-0003-3402-7212]{Swarnim Shashank}
\affiliation{Center for Astronomy and Astrophysics, Center for Field Theory and Particle Physics and Department of Physics, Fudan University, Shanghai 200438, China}
\email{swarnim@fudan.edu.cn, swarrnim@gmail.com}

\author[0000-0002-7671-6457]{Askar B. Abdikamalov}
\affiliation{School of Humanities and Natural Sciences, New Uzbekistan University, Tashkent 100001, Uzbekistan}
\affiliation{Ulugh Beg Astronomical Institute, Tashkent 100052, Uzbekistan}

\author[0000-0003-2845-1009]{Honghui Liu}
\affiliation{Institut f\"ur Astronomie und Astrophysik, Eberhard-Karls Universit\"at T\"ubingen, D-72076 T\"ubingen, Germany}
\affiliation{Center for Astronomy and Astrophysics, Center for Field Theory and Particle Physics and Department of Physics, Fudan University, Shanghai 200438, China}

\author{Temurbek Mirzaev}
\affiliation{Center for Astronomy and Astrophysics, Center for Field Theory and Particle Physics and Department of Physics, Fudan University, Shanghai 200438, China}
\affiliation{Institute for Advanced Studies, New Uzbekistan University, Movarounnahr str. 1, Tashkent 100007, Uzbekistan}
\affiliation{Ulugh Beg Astronomical Institute, Tashkent 100052, Uzbekistan}

\author[0000-0002-9639-4352]{Jiachen Jiang}
\affiliation{Department of Physics, University of Warwick, Gibbet Hill Road, Coventry CV4 7AL, UK}
\affiliation{Institute of Astronomy, University of Cambridge, Madingley Road, Cambridge CB3 0HA, UK}

\author[0000-0002-3180-9502]{Cosimo Bambi}
\affiliation{Center for Astronomy and Astrophysics, Center for Field Theory and Particle Physics and Department of Physics, Fudan University, Shanghai 200438, China}
\affiliation{School of Humanities and Natural Sciences, New Uzbekistan University, Tashkent 100001, Uzbekistan}

\author{Fergus Baker}
\affiliation{School of Physics, University of Bristol, Tyndall Avenue, Bristol BS8 1TH, UK}

\author{Andrew Young}
\affiliation{School of Physics, University of Bristol, Tyndall Avenue, Bristol BS8 1TH, UK}

\correspondingauthor{Cosimo Bambi} \email{bambi@fudan.edu.cn}

\begin{abstract}
Relativistic reflection is a common feature in the X-ray observations of accreting compact objects. We present \texttt{reflux}, a new X-ray reflection model for spectral analysis of super-Eddington sources. We develop two relativistic reflection frameworks for super-Eddington accretion: a slim-disk model that self-consistently accounts for disk thickening and self-shadowing, and an optically thick wind model that treats reflection off a funnel-shaped outflow. The slim-disk model offers a geometry where the inner disk thickness is proportional to radius, becoming thicker as the mass accretion rate increases. The wind model measures the opening angle of the funnel, the wind speed, and wind acceleration radius. The slim-disk profile reduces the brightness of the blue horn in the Fe\,K emission line for a fixed emissivity and significantly changes the intensity profile for a lamppost geometry. The wind model shows a blue-shifted iron line due to high velocity outflows. Both models assume a spherically symmetric spacetime. We apply the wind model to the \textit{XMM-Newton} spectrum of the tidal disruption event Swift~J1644+57, where the Fe\,K profile is expected to be shaped by scattering in an outflowing funnel. We constrain the opening angle of the funnel and find a high velocity of the wind.
\end{abstract}

\keywords{Accretion (14) --- Ultraluminous x-ray sources (2164) --- X-ray astronomy (1810)}

\section{Introduction} \label{sec:intro}

Super-Eddington accretion has become an important phenomenon to be studied in the recent years in sources where the luminosity exceeds the Eddington luminosity ($L_{\rm Edd} \approx 1.26 \times 10^{38} M/M_{\odot} ~\mathrm{erg~s}^{-1}$). It has been observed in Ultraluminous X-ray sources (ULXs) \citep{Kaaret:2017tcn, 2021AstBu..76....6F} which are compact objects observed in nearby galaxies and even one Galactic pulsar Swift~J0243.6+6124 \citep{2017GCN.21960....1C, 2017ATel10809....1K}. Super-Eddington accretion has been observed in tidal disruption events (TDEs), which are transient events occurring when a star is disrupted \citep[see][for a recent review on TDEs]{Gezari:2021bmb} and accreted by a massive black hole \citep[see][]{2011Sci...333..203B, 2011Natur.476..421B, Kara:2016kbu, Kara:2017wle, Lin:2015yka}. There are also observations of super-Eddington accretion in active galactic nuclei (AGNs) \citep{Lanzuisi:2016gts, 2018ApJ...868...15T} where accretion takes place onto supermassive black holes at the centers of galaxies. Accretion beyond the critical Eddington limit is also considered an important factor of growth of massive black holes discovered in the early Universe \citep{Mortlock:2011va, 2023ARA&A..61..373F, 2023A&A...670A.180M, Matthee:2023utn}.

Fluorescent iron lines are commonly observed in the X-ray spectra of a number of astrophysical sources, such as neutron star \citep{2007ApJ...664L.103B,2008ApJ...674..415C} and black hole \citep{2021ApJ...913...79T,2023ApJ...946...19D,Liu2023ApJ...950....5L,2023ApJ...951..145L} X-ray binaries (XRBs), AGNs \citep{2007MNRAS.382..194N,2013MNRAS.428.2901W}, and TDEs \citep{2022ApJ...937....8Y,Kara:2016kbu}. The source of this emission line is predicted to be the ``reflection" taking place at the accretion disk (or some optically thick medium like winds). The Comptonized photons from hot plasma around the central compact object interact with the material in the disk or outflow. This imprints the atomic data of the gas as emission and absorption lines onto the spectrum. However, due to the compact object there is strong gravity which blurs this reflection spectrum arising from gravitational redshift and Doppler boosting \citep[see, e.g.,][]{2017bhlt.book.....B,1995Natur.375..659T,Fabian:1989ej}. The relativistic reflection spectrum is characterized by the Compton hump and blurred elemental emission lines, out of which Fe\,K emission line at 6.4-6.9 keV is often the strongest feature \citep{2005MNRAS.358..211R}. Studying these reflection features is termed as X-ray reflection spectroscopy \citep[see, e.g.,][]{2021SSRv..217...65B}: it provides a useful tool to probe the accretion dynamics and gravitational effects near to the compact object and remains one of the leading techniques to measure black hole spins \citep{2006ApJ...652.1028B,2023ApJ...946...19D} and even performing tests of fundamental physics \citep{2018PhRvL.120e1101C,Tripathi:2019bya,2019ApJ...875...56T}. 

Current reflection models are designed to work in the thin-disk regime, which is normally thought to hold up to 20-30\% of the Eddington accretion rate \citep{Shakura:1972te,1973blho.conf..343N}. At higher accretion rates, the assumptions of the standard thin‐disk model begin to break down. When a source exceeds its Eddington limit, radiative diffusion becomes inefficient and a non‐negligible fraction of the dissipated energy is advected inward rather than being locally radiated \citep{Abramowicz:1988sp}. In this super-Eddington regime, the accretion flow will instead adopt a slim-disk structure and the disk no longer remains geometrically thin, in which radial advection and pressure gradients play a dominant role \citep{Abramowicz:1988sp,Watarai:2006gv,Sadowski:2009gg}. More recently, from the studies using radiative general relativistic magnetohydrodynamic simulations of super-Eddington accretion flows, it is also predicted that fast and optically think winds can be launched from the super-Eddington system due to magnetic and radiation pressures \citep{McKinney:2013txa,Jiang:2014tpa,Jiang:2017mbm,Sadowski:2014awa,Sadowski:2015hia,Thomsen:2022voj}. This optically thick wind creates a funnel-like geometry in which reflection happens (see e.g. Fig.~\ref{fig:wind}).

However, the systematic effects introduced by the common assumption of a razor-thin accretion disk geometry \citep{1973blho.conf..343N, Page:1974he} have not been thoroughly investigated. Some studies \citep[e.g.,][]{Shashank:2022xyh, Jiang:2022sqv, Tripathi:2021wap, Abdikamalov:2020oci, Taylor:2017jep} have explored this topic, but they remain within the geometrically thin regime suitable for sub-Eddington accretion disks ($H/r << 1$). \citet{Jiang:2022sqv} found that the systematic uncertainty is small compared to the statistical uncertainty of current data. \citet{Shashank:2022xyh,Shashank:2025hka} studied these systematic uncertainties by fitting the data with the current reflection models to spectra obtained from numerically simulated disks and found that for current missions (\textit{NuSTAR}) the models work well for fast spinning black holes. Effects of geometrically thick disks were explored in \citet{Riaz:2019bkv, Riaz:2019kat}.  

Future high-resolution data from missions like eXTP, AXIS, or NewAthena will require more precise models. This is particularly important in the super-Eddington regime, as more objects such as TDEs and ULXs exhibit relativistic reflection spectra in their X-ray observations \citep[e.g.,][]{Jaisawal2019ApJ...885...18J, 2022ApJ...937....8Y, Yao:2024yrc, Kara:2016kbu, Motta:2017mol, Veledina:2023zho}. A full model which takes into account slim disks and optically thick winds for super-Eddington scenarios have not been developed for studying reflection spectra (as per the authors' knowledge). We note that there are indeed existing models for the thermal emissions arising from slim-disk geometries \citep{Kawaguchi:2003jz, Sadowski:2011vs, Caballero-Garcia:2017oqv, Wen:2022mvr}. 
Recently, \citet{Zhang:2024pzd} studied iron line emissions modeled from the optically thick winds in a super-Eddington regime.

In this work, we introduce \texttt{reflux}, a new relativistic reflection model with two geometries: $(i)$ slim-disk profile for super-Eddington accretion following \citet{Watarai:2006gv} and $(ii)$ optically thick wind model similar to \citet{Zhang:2024pzd}. 

The article is organized as follows. In Section~\ref{sec:model}, we describe the model {\tt reflux}. Section~\ref{sec:fit} shows working of the model by fitting the 2011 \textit{XMM-Newton} observation of TDE Swift~J1644+57. Finally, we discuss the results and future directions in Section~\ref{sec:discussion}. Units used in this article are $G=c=1$ unless mentioned otherwise.

\section{Model description} \label{sec:model}

To construct a full relativistic reflection model, we assume Schwarzschild spacetime around the central object. Further we use the same methods of ray-tracing \citep{raytransfer} and Cunningham's transfer function \citep{Cunningham:1975zz, 1995CoPhC..88..109S,Dauser:2010ne} as demonstrated in detail in \citet{Bambi:2016sac, Abdikamalov:2019yrr, Abdikamalov:2020oci} \citep[specifically, exactly the same methods have been used to calculate the transfer functions as discussed in Sec. 3 of][]{Abdikamalov:2020oci}. The new parameters in \texttt{reflux} related to the super-Eddington accretion geometries are discussed in the next sections. The parameters related to \texttt{xillver} tables are not discussed here as they are same as \texttt{relxill}\footnote{The parameters of the \texttt{relxill} and \texttt{xillver} model can be found here: \url{https://www.sternwarte.uni-erlangen.de/~dauser/research/relxill/index.html}.}.

\subsection{Slim-disk}

\begin{figure}[htpb]
    \centering
    \includegraphics[width=0.5\textwidth,trim=2.5cm 0.5cm 2.5cm 0.0cm,clip]{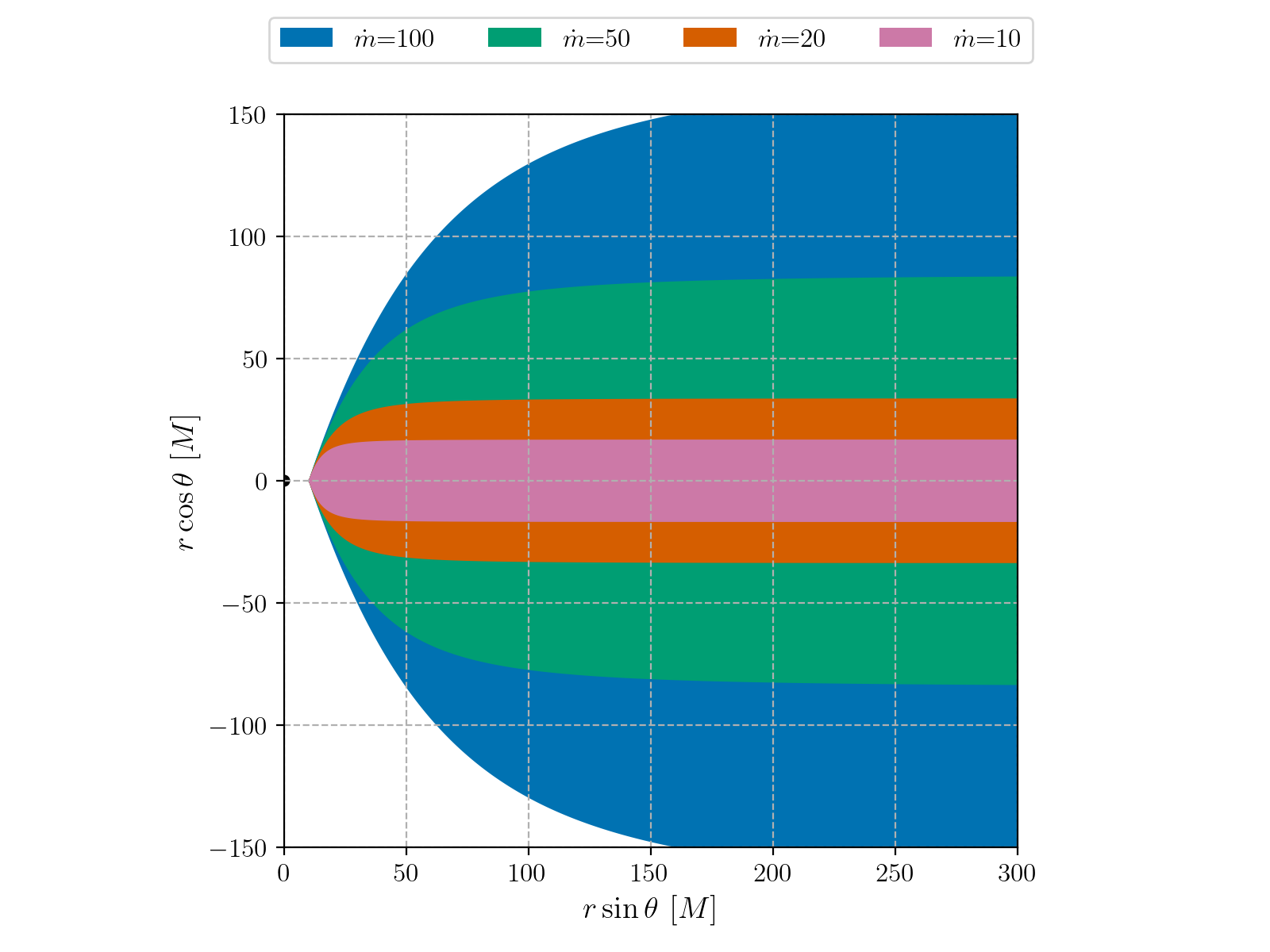}
    \caption{Disk profile obtained from the model for various accretion rates. The accretion rate $\dot{m}$ is in units of $\dot{M}_{\rm crit}$. The inner edge is set to $10M$. The black region at the origin denotes the event horizon of a Schwarzschild black hole.}
    \label{fig:disk_profile}
\end{figure}

To model the expected slim-disk profile for a supercritical accreting source, we use the analytical disk model introduced by \citet{Watarai:2006gv}. The model assumes an optically thick disk by introducing a radially dependent ratio $f = \mathcal{Q}^-_{\rm adv}/\mathcal{Q}^+_{\rm vis}$. $\mathcal{Q}^-_{\rm adv}$ is the advective cooling rate and $\mathcal{Q}^+_{\rm vis}$ is the viscous heating rate \citep[for their explicit forms, see][]{Watarai:2006gv}. $f$ also depends on the accretion rate, which then controls the height of the disk in our implementation. The scale height of the disk is defined as
\begin{equation}
    H = \left[ (2N+3) \frac{B \Gamma_{\Omega} \Omega_{0}^{2}}{\xi} \right]^{1/2} f^{1/2} \tilde{r} \, ,
\end{equation}
where $\Omega_{0}$ is a multiplier to the Keplerian angular velocity ($\Omega = \Omega_{0} \Omega_{\rm K}$), which we set to unity for having a disk with Keplerian velocity ($\Omega_{\rm K}$), and $\Gamma_{\Omega} = -d \log \Omega/d \log r$ is the linear approximation form of the angular velocity. We take $B=1$, $N=3$, $\xi=1.5$, and $\Gamma_{\Omega}=1.5$ as detailed in \citet{Watarai:2006gv}. $f$ is defined as
\begin{equation}
    f(x) = \frac{1}{2} \left[D^2 x^2 + 2 - D x (D^2 x^2 + 4)^{1/2} \right] \, ,
\end{equation}
where
\begin{equation}
    D = \left[\frac{64 \Omega_{0}^2}{(2N+3) \xi B \Gamma_{\Omega}}\right]^{1/2}, \quad x = \frac{\tilde{r}}{\dot{m}} \, ,
\end{equation}
$\dot{m} = \dot{M}/\dot{M}_{\rm crit}$ is the accretion rate in units of Eddington accretion rate ($\dot{M}_{\rm crit} = L_{\rm Edd}$, where $L_{\rm Edd}$ is the Eddington luminosity of the source), and $\tilde{r}$ is the disk radius. In our implementation of the model, we only vary $\dot{m}$, which gives us a family of disks whose scale heights are controlled by the value of $\dot{m}$. Also, we keep the inner edge of the disk as a free parameter, hence the $\tilde{r}$ represented above is the disk radius starting from the inner edge. Fig.~\ref{fig:disk_profile} shows the disk profile obtained by changing the value of the accretion rate.

\begin{figure*}[htpb]
    \centering
    \includegraphics[width=0.95\textwidth]{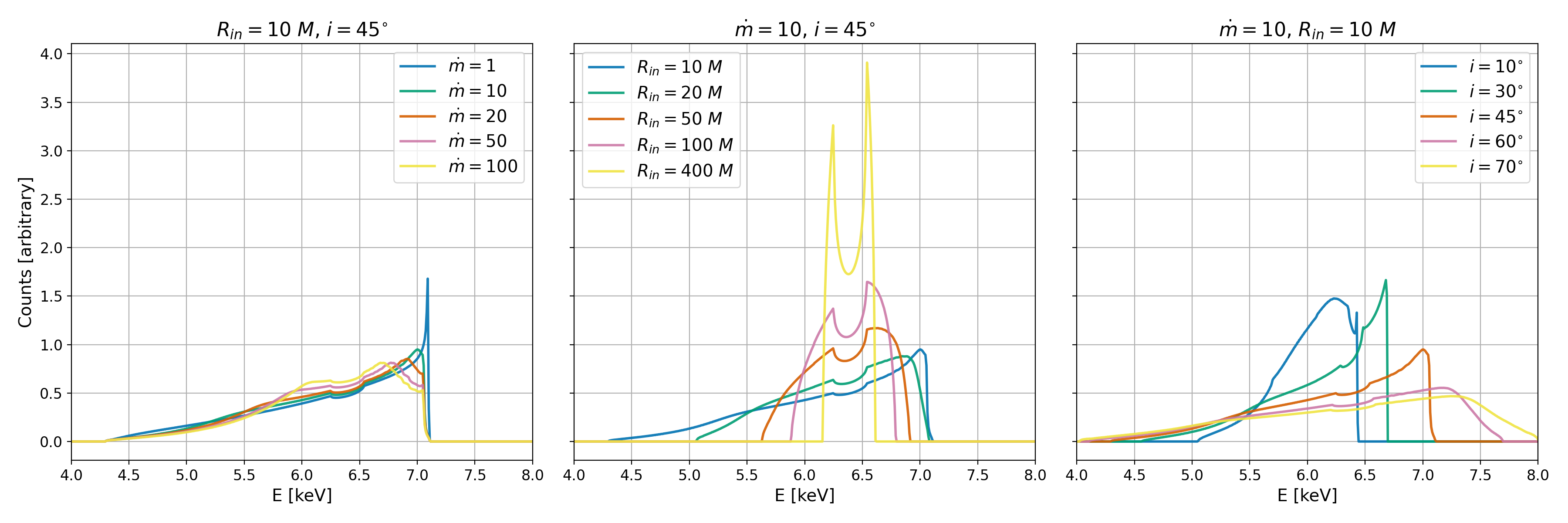}
    \caption{Iron lines obtained from the model for different parameter values. The left panel shows some iron line profiles for different values of the mass accretion rate, the central panel shows some iron line profiles for different values of inner edge of the disk, and the right panel shows some iron line profiles for different values of the disk inclination angle. The other fixed parameter values are on top of every panel. The emissivity index is set to $q_{in}=q_{out}=3$ in all cases.}
    \label{fig:ironlines}
\end{figure*}

\begin{figure*}[htpb]
    \centering
    \includegraphics[width=0.95\textwidth]{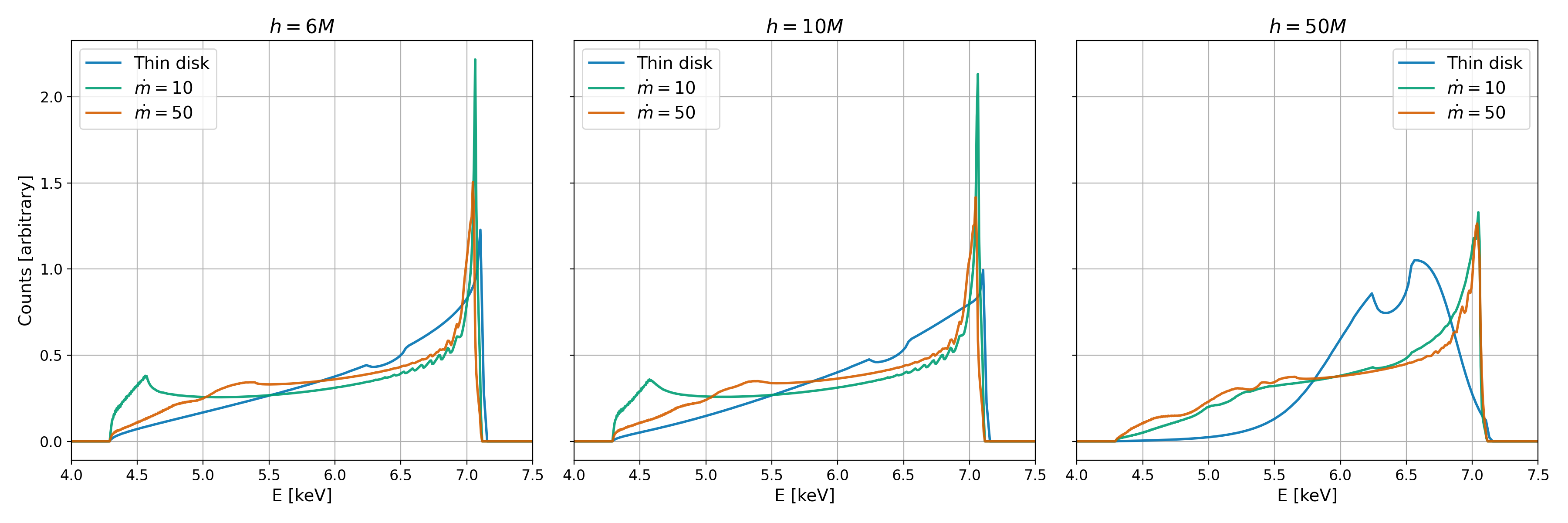}
    \caption{Iron lines obtained from the model for different disk heights using the lamppost emissivity. The left panel shows some iron line profiles for corona height $h=6M$, the central panel shows iron line profiles for $h=10M$, and the right panel shows iron line profiles for $h=50M$. The other fixed parameter values are $i=45^{\circ}$, $R_{in}=10M$ and $\Gamma=2$}
    \label{fig:ironlines_lp}
\end{figure*}

The free parameters of the model are: the inner edge of the disk ($R_{\rm in}$), which in the current version of {\tt reflux} varies from $6 M$ to $500 M$, the accretion rate ($\dot{m}$), which is varied from $1 \dot{M}_{\rm crit}$ to $100 \dot{M}_{\rm crit}$, and the inclination of the disk ($i$), varying from $3^{\circ}$ to $70^{\circ}$. While constructing the model, we also have the outer radius of the disk ($R_{\rm out}$) up to $1000 M$. For the lamppost version, we have the height of the lamppost corona ($h$) ranging from $2.2 M$ to $100 M$. With higher values of inclination, the photons from the inner regions of the disk do not reach the distant observer, so we do not consider inclinations higher than $70^{\circ}$. In Fig.~\ref{fig:ironlines}, we show the variations in iron line profiles for different parameter values: we see that the thickness of the disk produces a change in the shape of the iron lines, reducing the blue horn from thinner to a thicker disk. Fig.~\ref{fig:ironlines_lp} shows the iron line profiles for a lamppost emissivity. In the lamppost corona geometry, the emissivity profile of the disk significantly changes as the inner region of the disk is more illuminated due to the thicker disk. For low corona heights and high accretion rates, the outer regions of the disk are shadowed due to the thickness of the disk.

\subsection{Optically thick wind}

\begin{figure}[htpb]
\vspace{1.0cm}
    \centering
    \includegraphics[width=0.9\linewidth]{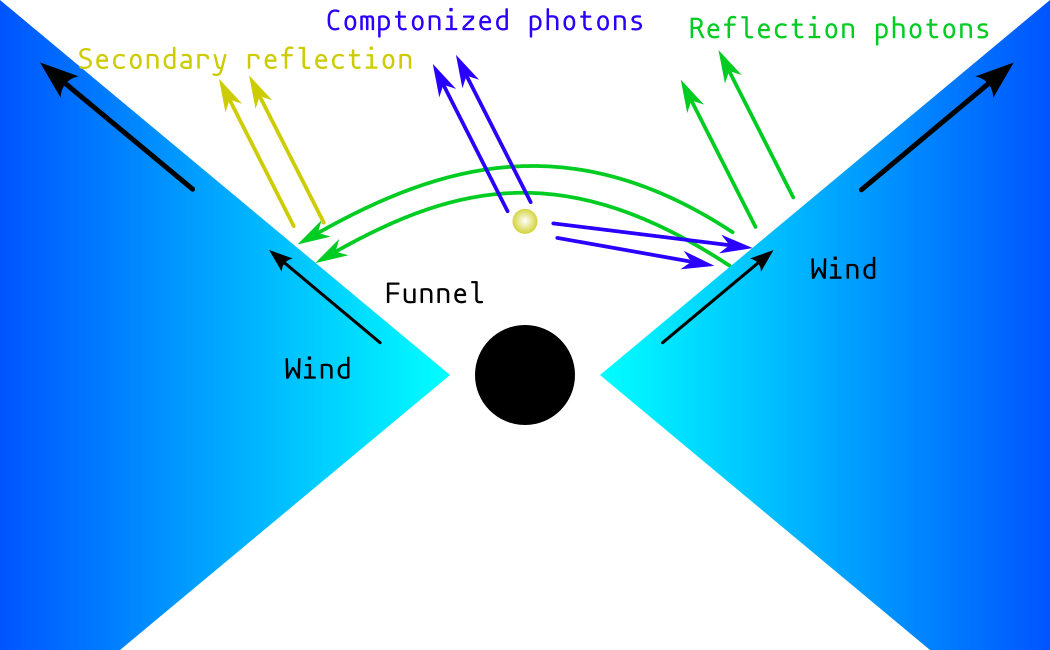}\\
    \vspace{0.2cm}
    \caption{Sketch depicting the geometry of the optically thick wind used for reflection modeling.}
    \label{fig:wind}
\end{figure}

To model the optically thick winds from a super-Eddington accreting system, we follow \citet{Zhang:2024pzd}. The winds launched from the system form a funnel structure within which the reflection takes place. Fig.~\ref{fig:wind} shows the geometry of the wind used to model the optically thick winds which produce reflection features. A hot and compact corona produces the power-law photons (blue arrows) which then reflect from the winds (green arrows), then these reflected photons return back to the reflection surface and undergo secondary reflection (yellow arrows). The funnel follows the velocity profile of the extended CAK law \citep{1975ApJ...195..157C,Thomsen:2019onb,Parkinson:2022ygo}
\begin{equation}
    v = v_0 + (v_{\infty} - v_0) \frac{[(r-r_{\rm ISCO})/R_{\rm acc}]^{\alpha}}{[(r-r_{\rm ISCO})/R_{\rm acc}]^{\alpha}+1} \, ,
\end{equation}
where $v_0$ is the wind launch velocity, which we take as $0$, and $v_{\infty}$ is the terminal velocity. $R_{\rm acc}$ is the wind acceleration radius. $r_{\rm ISCO}$ is the radius of the innermost stable circular orbit, which, in our case, is set to $6M$. The model has two emissivity profiles: broken power-law and lamppost.

\begin{figure*}[htpb]
    \centering
    \includegraphics[height=0.3\linewidth]{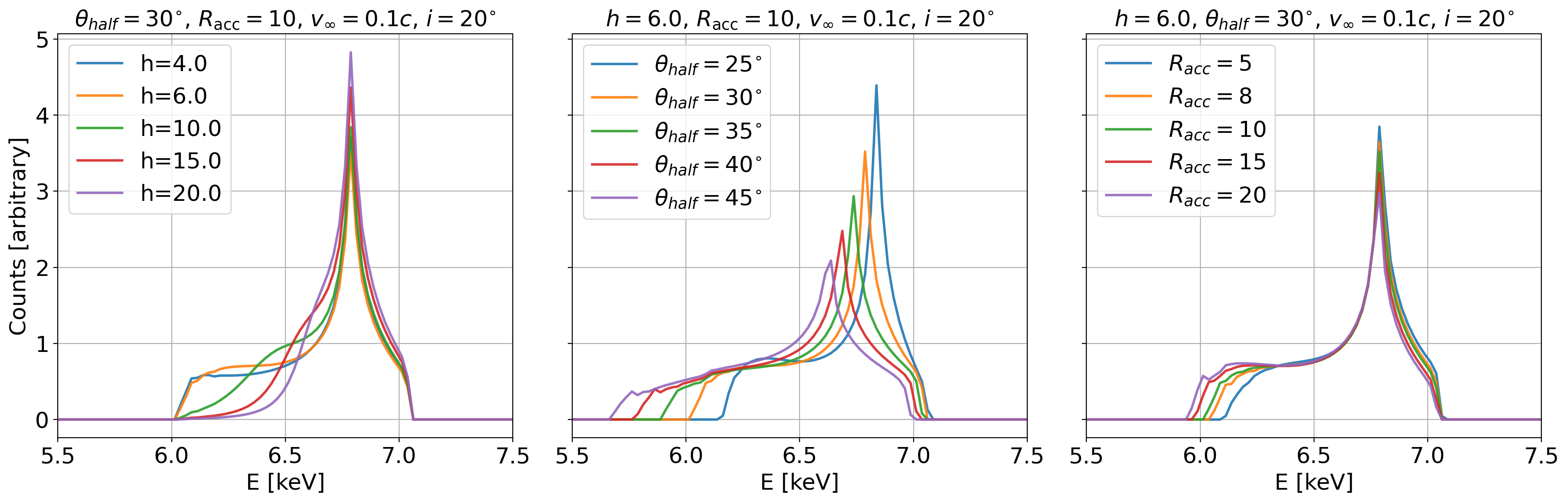}
    \includegraphics[height=0.3\linewidth]{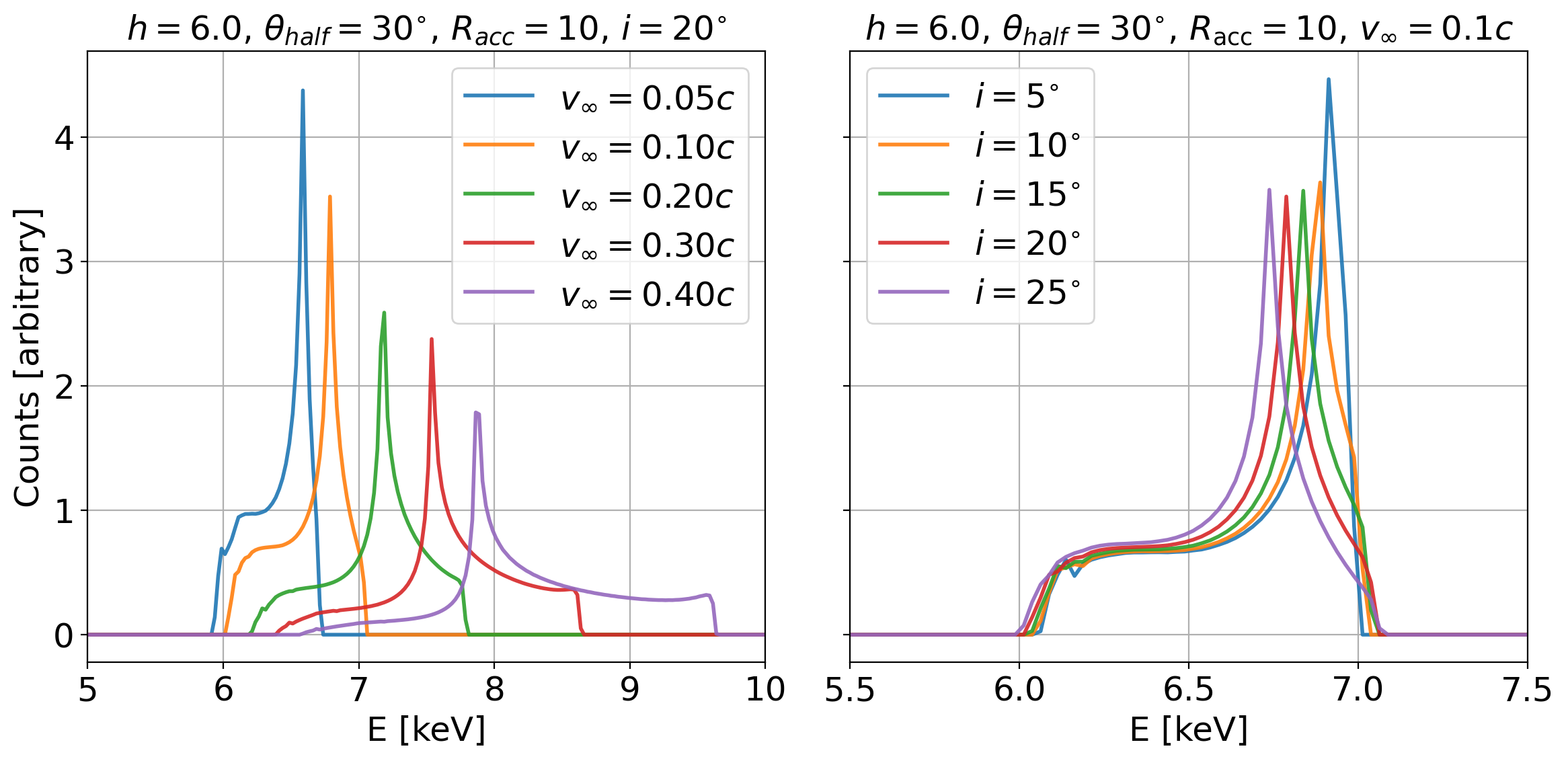}
    \caption{Iron line profiles for the wind reflection with a lamppost. Each panel varies one parameter while others are fixed. Top-left: dependence on lamppost height ($h$). Top-middle: dependence on wind opening angle ($\theta_{\rm half}$). Top-right: dependence on acceleration radius ($R_{\rm acc}$). Bottom-left: dependence on terminal wind speed ($v_{\infty}$). Bottom-right: dependence on inclination angle ($i$). The parameters are fixed as follows: $\Gamma=2$, $h=6M$, $\theta_{\rm half}=30^\circ$, $R_{\rm acc}=10M$, $v_{\infty}=0.1c$, and $i=20^\circ$.}
    \label{fig:wind_ironlines}
\end{figure*}

As discussed in \citet{Zhang:2024pzd}, the secondary reflection becomes important in a wind like geometry (Fig.~\ref{fig:wind}, yellow arrows), hence, in our lamppost models we use \texttt{ziji} \citep{Mirzaev:2024fgd,Mirzaev:2024qcu} to calculate the correct emissivity profiles for returning radiation and implement it in the model. This is similar to the returning radiation in the lamppost models of \texttt{relxill} \citep{Dauser:2022zwc}. The free parameters and their range in the model are follows. The half of opening angle of the wind funnel $\theta_{\rm half}$ is in the range $18^{\circ}$ to $45^{\circ}$. Acceleration radius ($R_{\rm acc}$) varies from $2M$ to $20M$. Terminal wind speed ($v_{\infty}$) is in the range $0$ to $0.5c$. The inclination angle ($i$) has lowest value $3^{\circ}$ and maximum is always equal to $\theta_{\rm half}$. In the lamppost model the height of the corona $h$ varies from $4M$ to $20M$. We also have the outer radius ($R_{\rm out}$) of the accretion flow up to $400M$.
Fig.~\ref{fig:wind_ironlines} shows the dependence of the iron lines on different parameter values in the model.

\section{Super-Eddington TDE Swift~J1644+57} \label{sec:fit}

In this section, we discuss the fitting of for TDE Swift~J1644+57 which was first discovered on 28 March 2011 by \textit{Swift} Burst Alert Telescope \citep{2011Natur.476..421B} at the center of a compact, non-interacting, star-forming galaxy with a redshift of $z=0.3534$. The source reached a peak luminosity of $\sim 10^{48}$~erg~s$^{-1}$. Follow up studies were done by \textit{Suzaku} and \textit{XMM-Newton} instruments \citep{Reis:2012sz}. Here, we re-analyze the \textit{XMM-Newton} observation (ObsID 0678380101, 16 March 2011).
The \textit{XMM-Newton} data were reduced using the Science Analysis System (SAS version 19.1.0) and the calibration files as of 2023. For the EPIC-pn camera, calibrated event files were produced with the EPPROC task. Background flaring was filtered by applying a threshold cut on the 10–12 keV light curve. The event list was then filtered using standard criteria \texttt{PATTERN <= 4} and \texttt{FLAG == 0}. The source spectrum was extracted from an annular region with an outer radius of 60 arcseconds, excluding a central circular region of 12 arcseconds to mitigate pile-up. The background spectrum was extracted from a circular region of the same size (60 arcseconds) located near the source. The redistribution matrix file (RMF) and ancillary response file (ARF) were generated using the RMFGEN and ARFGEN tasks, respectively.

\begin{figure*}[htpb]
    \centering
    \includegraphics[height=0.35\linewidth]{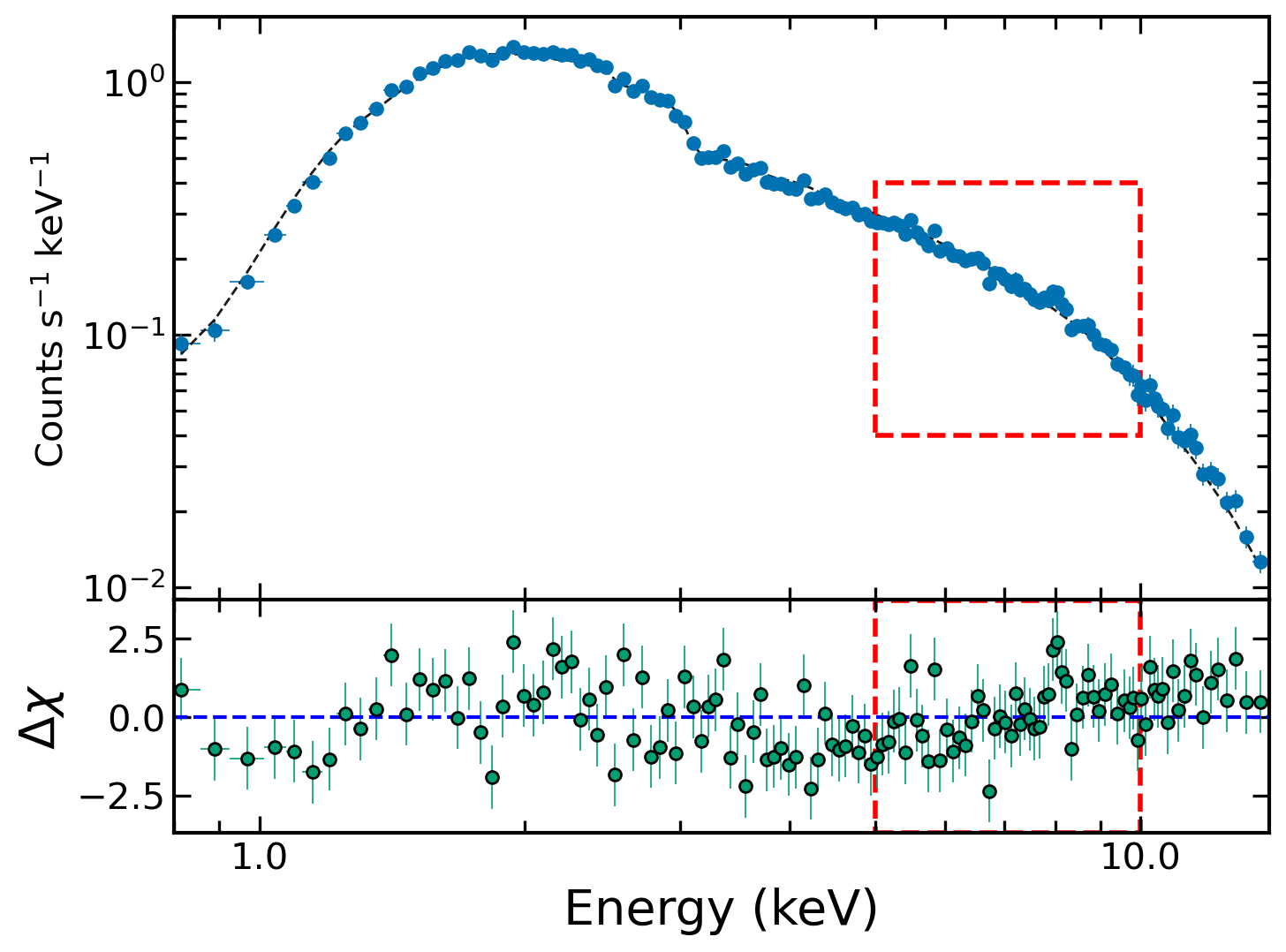}
    \includegraphics[height=0.35\linewidth]{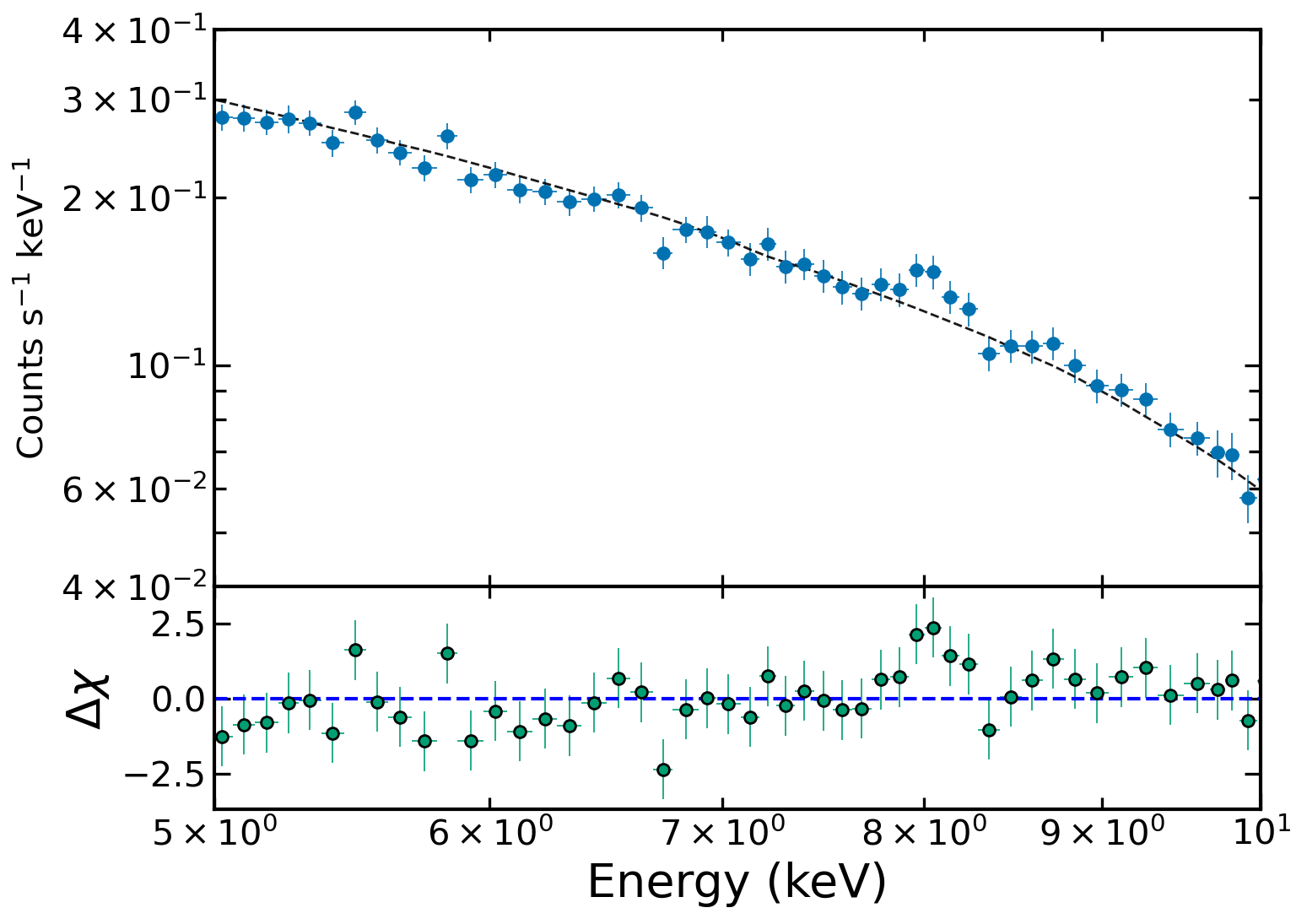}
    \caption{Swift J1644+57 data fitted with \texttt{TBabs*zTBabs*zcutoffpl}. Both panels show the same data; the right panel is zoomed in 5-10 keV. The red dashed boxes on the left panel denote the zoomed region.}
    \label{fig:tde-data-delchi}
\end{figure*}

We first fit the data with \texttt{TBabs*zTBabs*cutoffpl} with redshift set to 0.3534. We use \texttt{TBabs} to take into account galactic absorption and fix the value of $N_H$ to 0.0173\footnote{\url{https://www.swift.ac.uk/analysis/nhtot/index.php}}. Fig.~\ref{fig:tde-data-delchi} shows the data and the iron line obtained $\sim8$~keV, as also noted in \cite{Kara:2016kbu}. It was speculated from the energy and reverberation spectra that the source has a wind-funnel geometry \citep{Kara:2016kbu}, hence a good choice for fitting using the \texttt{reflux\_wind} model.

\begin{figure*}[htpb]
    \centering
    \includegraphics[width=0.32\linewidth]{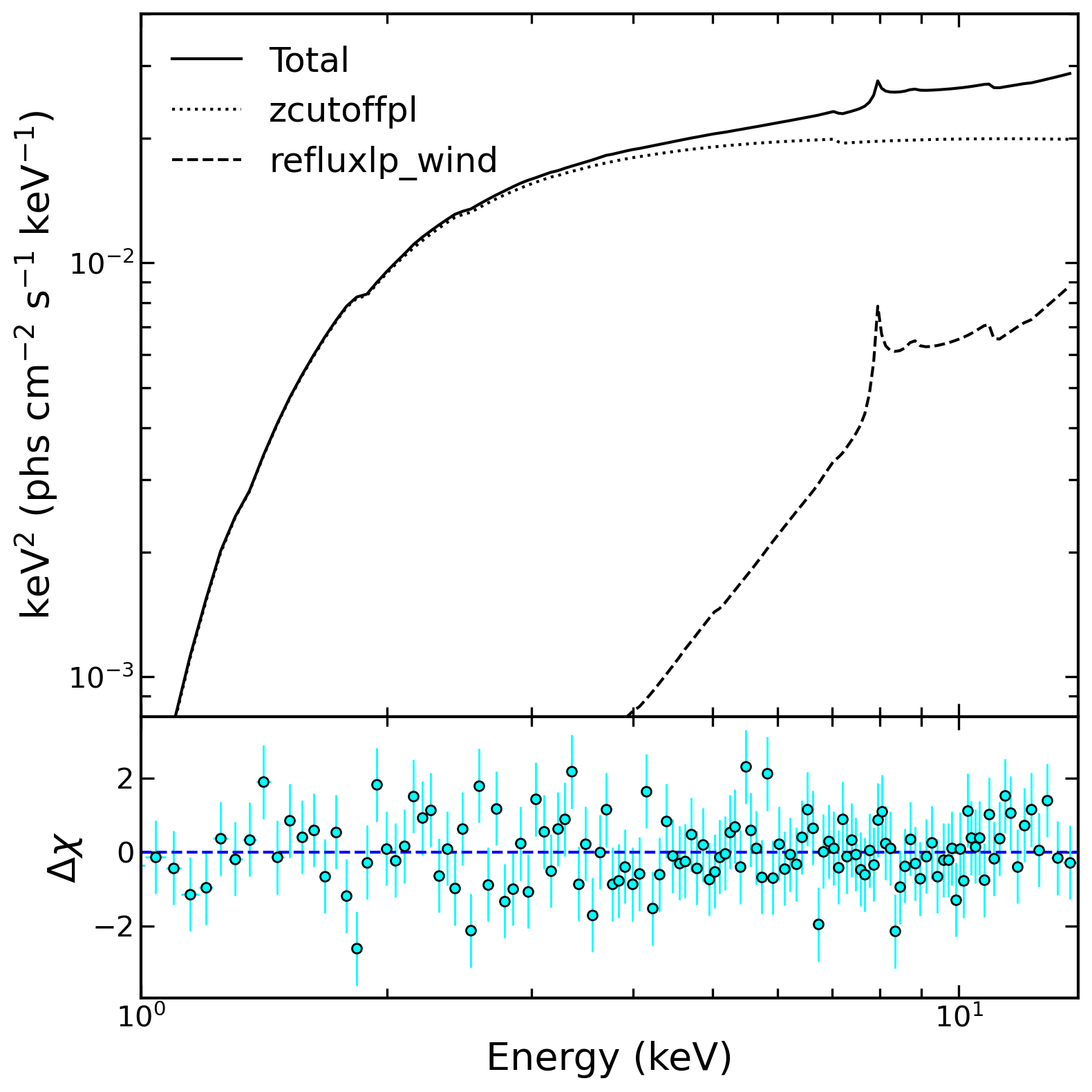}
    \includegraphics[width=0.32\linewidth]{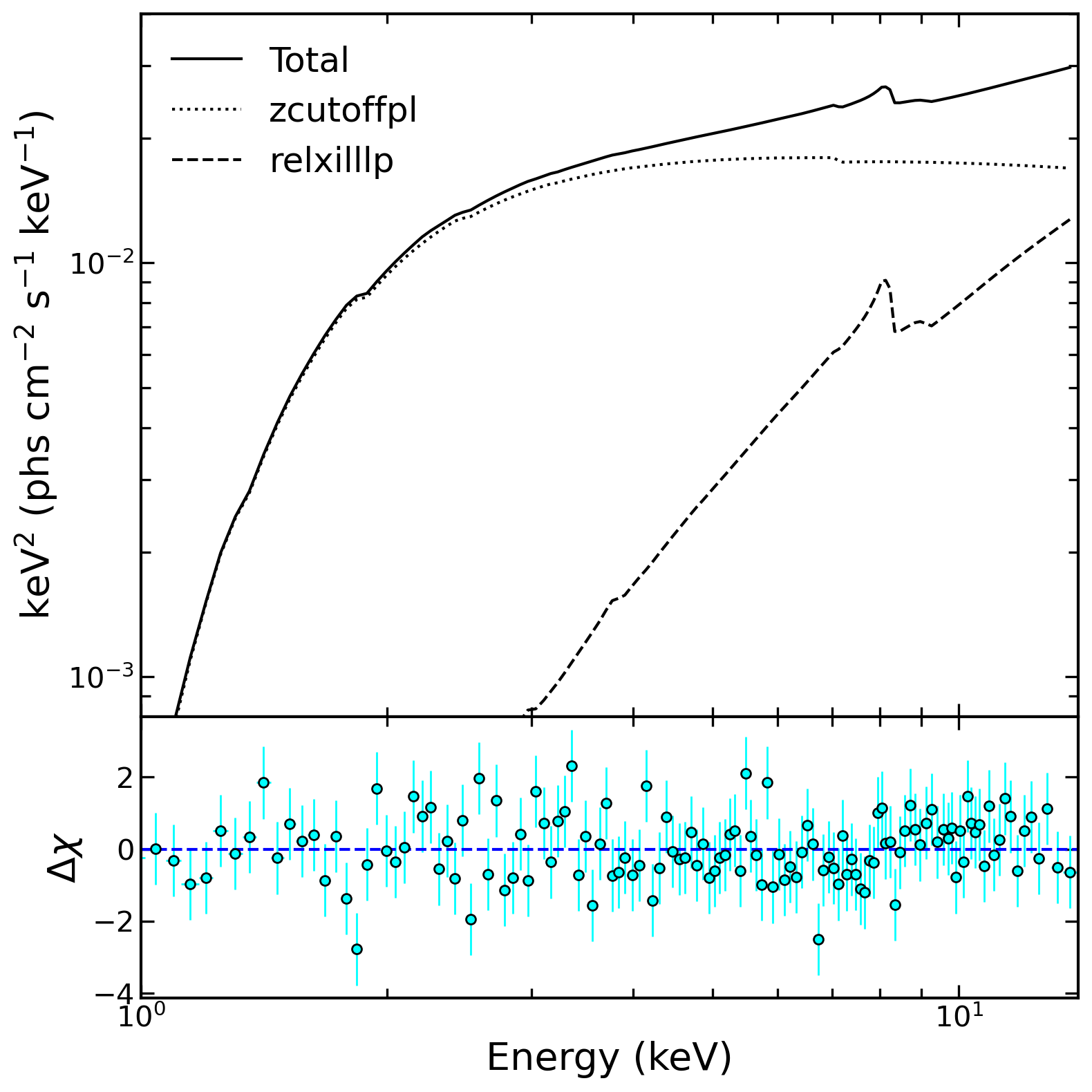}
    \includegraphics[width=0.32\linewidth]{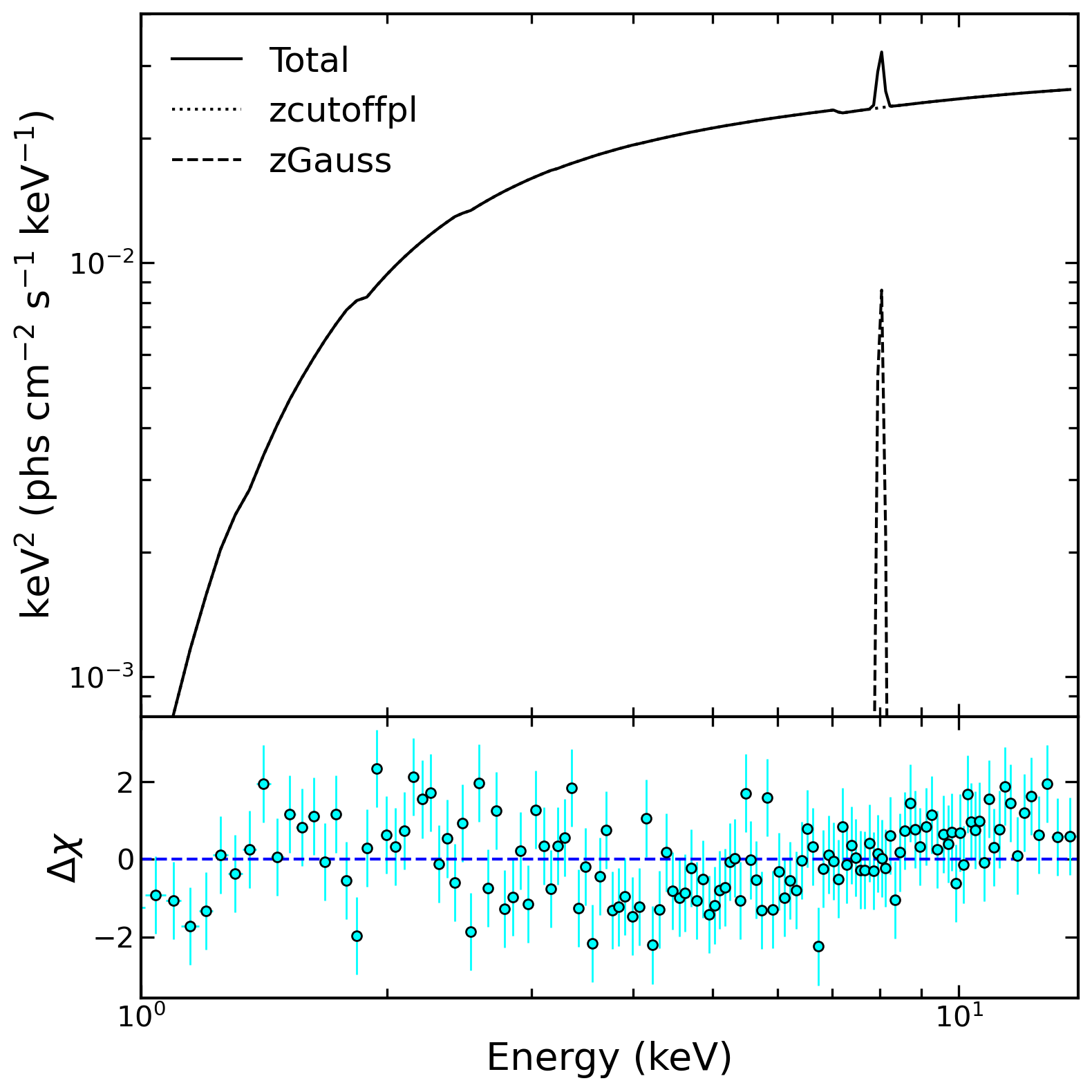}
    \caption{Best-fit model components and residuals for Swift~J1644+57 fitted with \texttt{reflux\_wind} (left), \texttt{relxilllp} (middle) and \texttt{zGauss} (right).}
    \label{fig:tde-fit}
\end{figure*}

To fit the reflection/iron line component, we use the model: \texttt{TBabs*zTBabs*(zcutoffpl+reflection)}. We perform three different fits with \texttt{reflection} set as \texttt{zGauss} for fitting only the iron line, \texttt{relxilllp} for testing with the thin-disk geometry and \texttt{refluxlp\_wind} to fit the wind funnel geometry. The best-fit results are shown in Tab.~\ref{tab:tde_fit} and Fig.~\ref{fig:tde-fit}.

\begin{table*}[htpb]
\centering
\small
\begin{tabular}{llllll}
\toprule
Model & Parameter & Unit & \texttt{refluxlp\_wind} & \texttt{relxilllp} & \texttt{zGauss} \\
\hline
\texttt{TBabs} & $N_{H}$ & $10^{22}$ cm$^{-2}$ & $0.0173^{*}$ & $0.0173^{*}$ & $0.0173^{*}$ \\
\texttt{zTBabs} & $N_{H}$ & $10^{22}$ cm$^{-2}$ & $1.557^{+0.049}_{-0.044}$ & $1.597^{+0.055}_{-0.059}$ & $1.443^{+0.029}_{-0.028}$ \\
\texttt{zcutoffpl} & $\Gamma$ &  & $1.923^{+0.042}_{-0.051}$ & $2.010^{+0.079}_{-0.087}$ & $1.748^{+0.020}_{-0.020}$ \\
 & $E_{\rm cut}$ & keV & $100^{*}$ & $100^{*}$ & $100^{*}$ \\
 & norm &  & $0.03440^{+0.0016}_{-0.0019}$ & $0.03680^{+0.0015}_{-0.0033}$ & $0.02874^{+0.0009}_{-0.0009}$ \\
\texttt{reflection} & $h$ & $M$ & $5.9^{+0.9}_{-1.5}$ & $2.00^{+0.003}_{-P}$ & -- \\
 & $a$ &  & -- & $0.801^{+0.148}_{-0.410}$ & -- \\
 & $\theta_{\rm half}$ & deg & $29.133^{+5.754}_{-4.434}$ & -- & -- \\
 & $v_{\infty}$ & $c$ & $>0.384$ & -- & -- \\
 & $R_{\rm acc}$ & $M$ & $10^{*}$ & -- & -- \\
 & $i$ & deg & $23.82^{P~\dagger}_{-8.71}$ & $68.0^{+P}_{-1.8}$ & -- \\
 & $\log\xi$ & erg cm s$^{-1}$ & $<0.51$ & $0.58^{+0.50}_{-P}$ & -- \\
 & $A_{\rm Fe}$ & solar & $<0.67$ & $0.50^{+0.08}_{-P}$ & -- \\
 & norm & $\times 10^{-5}$ & $1.294^{+6.503}_{-0.385}$ & $376.3^{+1160.0}_{-28.5}$ & -- \\
\texttt{Gaussian} & $E_{\rm line}$ & keV & -- & -- & $8.028^{+0.076}_{-0.078}$ \\
 & $\sigma$ & keV & -- & -- & $0.0595^{+0.127}_{-P}$ \\
 & norm & $\times 10^{-5}$ & -- & -- & $4.34^{+2.39}_{-2.02}$ \\
\hline\hline
 & $\chi^{2}/\nu$ & & $108.27/124$ & $111.82/125$ & $143.34/128$ \\
\hline
\end{tabular}
\caption{Redshift ($z$) in all models is set to $0.3534$. $^{*}$ denotes the parameter was frozen and $P$ denotes the parameter reached its lower/upper limit. The galactic hydrogen column density ($N_{H}$) in model \texttt{TBabs} was obtained from \url{https://www.swift.ac.uk/analysis/nhtot/index.php}. $^{\dagger}$The upper bound of inclination angle is always set to $\theta_{\rm half}$ in \texttt{refluxlp\_wind}. The reported uncertainties correspond to the 90\% interval limits for one relevant parameter ($\Delta\chi^2 = 2.71$).}
\label{tab:tde_fit}
\end{table*}

\section{Discussions} \label{sec:discussion}

We tested our wind-reflection framework on the super-Eddington TDE Swift~J1644+57. This source, powered by a tidal disruption of a star around a supermassive black hole, shows a highly super-Eddington accretion flow and has long been proposed as a case where powerful winds and relativistic jets coexist. The spectrum obtained with \textit{XMM-Newton} presents clear reflection signatures in the Fe\,K band that cannot be adequately described by a simple Gaussian or a thin-disk reflection. 

Our analysis with \texttt{refluxlp\_wind} suggests a wind funnel with an half-opening angle of $\sim 30^{\circ}$. The wind is found to be very fast, with terminal velocity $v_\infty\gtrsim0.384\,c$, and the system is observed at a low inclination $i\approx24^\circ$. The corona is compact, located at $h \sim 6~M$, which allows for effective illumination of the funnel walls while still remaining within the gravitational well of the black hole. These geometric and kinematic constraints together explain why the observed Fe\,K feature is moderately blueshifted (as can also be seen in Fig.~\ref{fig:wind_ironlines}). The photons are scattered and reprocessed by the optically thick wind before escaping and the line-of-sight is within the funnel region. The opening angle and wind velocity might be related to the accretion rate as suggested from numerical simulations in \citet{Thomsen:2019onb}. We are not able to constrain the wind acceleration radius ($R_{\rm acc}$), possibly due to weak dependence of the iron line on the parameter (Fig.~\ref{fig:wind_ironlines} top-right panel) and the data quality. Interestingly, we notice that the wind in non-ionized ($\log\xi\ \lesssim 0.51$) and has very low iron abundance ($A_{\rm Fe} \lesssim 0.67$). Just for comparison and testing, we also fit the data with \texttt{relxilllp} and a Gaussian. 

Out of the three, the lowest $\Delta\chi^2$ is obtained with \texttt{refluxlp\_wind}. The fitting results with \texttt{relxilllp} seem unphysical as the thin-disk geometry in the model predicts a high inclination of $i \sim 68^{\circ}$. At such a high inclination, the super-Eddington system should be obscured either by the slim-disk or the optically thick outflow. It also predicts extremely low corona height $h\sim2M$ and a spin of $a\sim0.8$. However, \texttt{relxilllp} fitting also predicts the low ionization and low iron abundance in the gas -- consistent with the wind model. The Gaussian line fitting reveals only a narrow line ($\sigma \sim 0.06$~keV) located at $\sim8$~keV with more residuals (Fig.~\ref{fig:tde-fit}). The fitting results of \texttt{zTBabs} and \texttt{zcutoffpl} are consistent across the three fitting results, with only the fit with \texttt{zGauss} having slight discrepancy (see Tab.~\ref{tab:tde_fit}).

From an astrophysical perspective, the wind geometry inferred for Swift~J1644+57 is consistent with hydrodynamic and magnetohydrodynamic (MHD) simulations of super-Eddington accretion, which predict fast, optically thick outflows launched from the inner disk. The funnel angle and velocity we measure are comparable to those suggested by radiation-MHD models.
A misclassification of wind reflection as disk reflection would bias spin or inclination estimates, highlighting the necessity of using models that capture the relevant geometry.

A more detailed analysis of the 2011 \textit{XMM-Newton} data of Swift~J1644+57 is beyond the purpose of the present work, which is the presentation of {\tt reflux}. The wind-reflection fit of Swift~J1644+57 emphasizes that super-Eddington systems can show fundamentally different reflection morphologies depending on whether the inner flow geometry is dominated by a geometrically thick disk or by a massive outflow. In all our fits, we have assumed that the spectrum illuminating the disk and responsible for the reflection component can be described by a power law with a high-energy cutoff. However, in the case of a TDE the reflection might be produced from the returning radiation of the thermal component from the disk, in such a case using {\tt xillverTDE} table might be useful \citep{Yao:2024yrc} but even better would be to have a reflection model which accounts for full returning radiation of thermal emissions \citep[e.g.][but with a super-Eddington geometry]{Mirzaev:2024qcu}. Nevertheless, the TDE Swift~J1644+57 was observed to have strong jet \citep{2011Natur.476..421B}, in such a case the base of the jet could act as the corona and our assumption stands well.

Our work motivates a broader application of both the slim-disk and wind-reflection models to a wide range of super-Eddington sources. These models applied with other spectral, timing and polarimetric data could really help decode the nature of accreting matter in extreme regimes. In future work, we plan to analyze data of sources which could possibly have the accretion geometry as in \texttt{reflux}. In terms of modeling, we wish to update \texttt{reflux} with the Kerr metric opening up the possibility of measuring black hole spins. We also hope to update the geometries with precise numerical accretion models.

\begin{acknowledgments}
We would like to thank the anonymous reviewer for their constructive comments to improve the manuscript content.
This work was supported by the National Natural Science Foundation of China (NSFC), Grant Nos.~12250610185, W2531002, and 12261131497.
S.S. acknowledges support from Shanghai Super Postdoctoral fellowship.
J.J. acknowledges support from Leverhulme Trust, Isaac Newton Trust and St Edmund's College, University of Cambridge.
Part of the computations in this research were performed using the CFFF platform of Fudan University.
\end{acknowledgments}

%\vspace{5mm}
\facilities{Swift, NICER, NuSTAR, XMM-Newton}

\software{astropy \citep{Astropy:2013muo,Astropy:2018wqo}, 
          XSPEC \citep{xspec}, 
          numpy \citep{numpy}, 
          scipy \citep{scipy}, 
          matplotlib \citep{matplotlib}, 
          raytransfer \citep{raytransfer}, 
          blackray \citep{blackray}, 
          blacklamp \citep{blacklamp},
          ziji \citep{mirzaev_2024_13954302} 
          }

\bibliography{references}{}
\bibliographystyle{aasjournal}

\end{document}